\documentclass[prd,aps,twocolumn,nofootinbib]{revtex4}

\usepackage{graphicx}

\begin{document}

\title{Baryogenesis from the Gauge-mediation type $Q$ ball \\ and the New type of $Q$ ball as dark matter}

\author{Shinta Kasuya$^a$ and Masahiro Kawasaki$^{b,c}$}

\affiliation{
$^a$ Department of Mathematics and Physics,
     Kanagawa University, Kanagawa 259-1293, Japan\\
$^b$ Institute for Cosmic Ray Research,
     the University of Tokyo, Chiba 277-8582, Japan\\
$^c$ Kavli Institute for the Physics and Mathematics of the Universe (WPI), 
     Todai Institutes for Advanced Study, the University of Tokyo, Chiba 277-8582, Japan}

\date{February 19, 2014}

\begin{abstract}
We investigate that the two types of the $Q$ balls explain the baryon asymmetry and
the dark matter of the universe in the gauge-mediated supersymmetry breaking. The gauge-mediation 
type $Q$ balls of one flat direction produce baryon asymmetry, while the new type $Q$ balls of another
flat direction become the dark matter. We show that the dark matter new type $Q$ balls are free from
the neutron star constraint. $n=5$ gauge mediation type and $n=6$ new type $Q$ balls are displayed
as an example, where the potential is lifted by the superpotential $\Phi^n$. 
These dark matter $Q$ balls may be detected by future observations, such as in advanced 
IceCube-like observations. 
\end{abstract}



\maketitle

\section{Introduction}
Affleck-Dine baryogenesis is one of the promising mechanism for creating baryon asymmetry of the 
universe \cite{AD}. It utilizes a scalar field called the Affleck-Dine (AD) field which carries baryonic charge 
for producing baryon number when it rotates in the potential. In suspersymmetry (SUSY), there are a lot of 
flat directions whose scalar potential vanishes in the SUSY limit. They consists of some combination of 
squarks (and sleptons) \cite{DRT, GKM}. The scalar potential is lifted by the SUSY breaking effects and 
the higher order operators coming from the nonrenormalizable superpotential. The field acquires a large field amplitude during inflation, and starts rotation 
when the Hubble parameter becomes the mass scale of the potential, $H\sim m_{\rm eff}$. In the end, the field 
decays into quarks to be the baryon number of the universe. 

Actually, the condensate of the AD field feels spatial instabilities during its rotation, and disintegrated into
spherical lumps, $Q$ balls \cite{KuSh}. The $Q$ ball is a nontopological soliton, the energy minimum 
configuration of a scalar field whose stability is guaranteed by a finite charge $Q$ \cite{Coleman}. 
The charge $Q$ is essentially the baryon number for the AD field. In gauge mediation, large enough $Q$ 
balls are stable against the decay into nucleons, the lightest particles with unit baryon number, since the 
mass per charge of the $Q$ ball is less than that of nucleons. In this case, the $Q$ ball can be the dark 
matter of the universe, while the baryons evaporated from the $Q$ balls would be the baryon number 
of the universe \cite {KuSh,KK3}, since all the baryonic charges are absorbed into formed $Q$ balls \cite{KK1,KK3}. 

On the other hand, the $Q$ ball may decay into baryons if the mass per charge of the $Q$ ball is larger than
that of the decay baryonic particles. This is achieved when the charge of the $Q$ ball is small enough. In this case,
the lightest SUSY particles (LSPs), such as the gravitino \cite{KK4,KKY, gravitino} or the axino\cite{KKK}, 
decayed from the $Q$ ball, may be the dark matter, while the baryons are produced by the $Q$-ball decay as well.

In the former case, however, the direct detection of the dark matter $Q$ balls and the estimation of the
cosmological parameters has been refined in this decade since our previous work \cite{KK3}, we can now
conclude that the dark matter $Q$ balls cannot explain the baryon asymmetry of the universe by those
baryonic charges evaporated from their surface. See the Appendix for more details. 

Then the next question is: Can we still explain the baryon number of the universe by the AD mechanism,
while the $Q$ ball is the dark matter? The answer is yes. What we show in this article is that the small enough
gauge-mediation type $Q$ balls \cite{KK3} of one flat direction decay into baryons, and the new 
type $Q$ balls \cite{new} of another flat direction are stable to be the dark matter of the universe. 
The peculiar feature of the dark matter new type $Q$ ball is that it does not suffer from the neutron star constraints,
raised in Ref.~\cite{NS}.  

The constructions of the article is as follows. In the next section, we review the features of the gauge-mediation
type and the new type $Q$ balls in the gauge mediation. The decay of the gauge-mediation type $Q$-ball is 
also explained in this section. Abundances of the baryon number and the dark matter in the universe is estimated 
in Sec.III. We show general constraints on both types of $Q$ balls in Sec.IV, and give an example of $n=5$ unstable 
gauge-mediation type and $n=6$ stable new type $Q$ balls in Sec.V. We conclude in Sec.VI. In the Appendix,
we show the observations exclude the scenario that the dark matter $Q$ ball with evaporated charge being
the baryon asymmetry of the universe.

\section{$Q$ ball in the gauge mediation}
The AD field is a flat direction which is some combination of squarks and sleptons\cite{DRT,GKM}. 
It has a large field value during and after inflation due to the balance between the negative 
Hubble-induced mass term, $V_H\simeq -H^2|\Phi|^2$, and the higher order term, 
$V_{\rm NR} \simeq |\Phi|^{2(n-1)}/M_{\rm P}^{2(n-3)}$, which stems from the nonrenormalizable 
superpotential, $W_{\rm NR} \simeq \Phi^n/M_{\rm P}^{n-3}$, where 
$M_{\rm P} (\approx 2.4 \times 10^{18}$ GeV) is the reduced Planck scale. When the the Hubble parameter
becomes comparable to the curvature scale of the potential, $H_{\rm rot} \simeq \sqrt{V''(\phi_{\rm rot})}$,
the field starts to oscillate with the helical motion caused by the $A$ term of the form, 
$V_A\simeq m_{3/2} W_{\rm NR}$, where $m_{3/2}$ is the gravitino mass.
During the rotation of the AD field, it feels spatial instabilities for the potential shallower than $\phi^2$.
Finally they grow into lumps, $Q$ balls.

In gauge-mediated SUSY breaking, the potential is lifted as \cite{flat, KuSh, EnMc}
\begin{equation}
V = V_{\rm gauge} + V_{\rm grav},
\end{equation}
where 
\begin{equation}
V_{\rm gauge}=\left\{\begin{array}{ll}
m_\phi^2 |\Phi|^2 & (|\Phi| \ll M_m), \\[2mm]
M_F^4 \left(\log\frac{|\Phi|^2}{M_m^2}\right)^2 & (|\Phi| \gg M_m),
\end{array}\right.
\label{pot-gauge}
\end{equation}
\begin{equation}
V_{\rm grav} = m_{3/2}^2|\Phi|^2\left( 1+K\log\frac{|\Phi|^2}{M_*^2}\right).
\end{equation}
Here, $m_\phi \sim O({\rm TeV})$ is a soft breaking mass scale, $M_m$ the messenger scale, 
$K(<0)$ a coefficient of one-loop correction, and $M_*$ the renormalization scale. 
Since the gravitino mass is much smaller than the weak scale, $V_{\rm grav}$ term dominates over 
$V_{\rm gauge}$ for larger field amplitudes. $M_F$ ranges as
\begin{equation}
10^3 \ {\rm GeV} \lesssim M_F \lesssim \frac{g^{1/2}}{4\pi}\sqrt{m_{3/2} M_{\rm P}}.
\label{MFlimit}
\end{equation}

There are two types of $Q$ balls in the gauge mediation: the gauge-mediation type \cite{KuSh} and 
the new type \cite{new}. In the former case, the potential is dominated by $V_{\rm gauge}$, and the 
$Q$ ball is created with the charge \cite{KK3}
\begin{equation}
\label{Qform}
Q_{\rm G}=\beta_{\rm G} \left(\frac{\phi_{\rm rot}}{M_F}\right)^4,
\label{form-gauge}
\end{equation}
where $\phi_{\rm rot}$ is the amplitude of the field at the onset of the rotation.
$\beta_{\rm G} \simeq 6\times 10^{-4}$ for a circular orbit ($\varepsilon=1$), while
$\beta_{\rm G} \simeq 6\times 10^{-5}$ for an oblate case ($\varepsilon\lesssim 0.1$). Here
$\varepsilon$ is the ellipticity of the field orbit. The properties of the $Q$ ball are given by
\begin{eqnarray}
&& M_Q \simeq \frac{4\sqrt{2}\pi}{3} \zeta M_F Q_{\rm G}^{3/4}, \\
&& R_Q \simeq \frac{1}{\sqrt{2}} \zeta^{-1} M_F^{-1} Q_{\rm G}^{1/4}, \\
&& \omega_Q \simeq \sqrt{2} \pi \zeta M_F Q_{\rm G}^{-1/4}, \\
&& \phi_Q \simeq \frac{1}{\sqrt{2}} \zeta M_F Q_{\rm G}^{1/4},
\end{eqnarray}
where $M_Q$ and $R_Q$ are the mass and  the size of the $Q$ ball, respectively, and $\omega_Q$ and 
$\phi_Q$ are respectively the rotation speed and the amplitude of the field inside the $Q$ ball, and 
$\zeta$ is the $O(1)$ parameter \cite{HNO, KY}.

On the other hand, for the latter case, where the potential is dominated by $V_{\rm grav}$, the $Q$ ball
with the charge $Q_N$ forms, where \cite{KK2, new}
\begin{equation}
Q_{\rm N} = \beta_{\rm N} \left(\frac{\phi_{\rm rot}}{m_{3/2}}\right)^2.
\label{form_new}
\end{equation}
Here $\beta_{\rm N}\simeq 0.02$ \cite{Hiramatsu}. The properties of the new type $Q$ ball are as follows: 
\begin{eqnarray}
&& M_Q \simeq m_{3/2} Q_{\rm N}, \\
&& R_Q \simeq |K|^{-1/2} m_{3/2}^{-1}, \\
&& \omega_Q \simeq m_{3/2}, \\
&& \phi_Q \simeq m_{3/2} Q_{\rm N}^{1/2}.
\end{eqnarray}
Actually, the charge $Q$ is the $\Phi$-number,
and relates to the baryon number of the $Q$ ball as
\begin{equation}
B=bQ,
\end{equation}
where $b$ is the baryon number of $\Phi$-particle. For example, $b=1/3$ for the $udd$ direction. 

The $Q$ ball is stable against the decay into nucleons for large field amplitude when the charge
is very large. The stability condition is given by $\omega_Q < m_{\rm D}$ where $m_{\rm D}$ is the 
mass of the decay particles. It generically holds that the new type Q ball is 
stable against the decay into nucleons, except for the gravitino mass larger than that of nucleons.
On the other hand, for the gauge-mediation type $Q$ ball, it reads as $Q_{\rm G} > Q_{\rm D}$,
where 
\begin{eqnarray}
\hspace{-5mm}
Q_{\rm D} & \equiv & 4\pi^4 \zeta^4 \left(\frac{M_F}{bm_N}\right)^4, \nonumber \\
& \simeq & 1.2 \times 10^{30} \left(\frac{\zeta}{2.5}\right)^4 \left(\frac{b}{1/3}\right)^{-4}
\left(\frac{M_F}{10^6 \ {\rm GeV}}\right)^4,
\label{QD}
\end{eqnarray}
with $m_N$ being the nucleon mass. Since we need unstable $Q$ ball to produce the baryon 
number of the universe, we consider the gauge-mediation type $Q$ ball with the charge smaller than 
$Q_{\rm D}$.

The unstable gauge-mediation type $Q$ ball decays into baryons through its surface~\cite{Cohen}~\footnote{
The abundance of NLSPs produced by the $Q$-ball decay is many orders of magnitude smaller than 
the BBN bound in our successful scenario.
}. 
The decay rate is refined in Refs.\cite{KY,KKY} as
\begin{equation}
\label{decayquark}
\Gamma_Q  \simeq  N_q \frac{1}{Q} \frac{\omega_Q^3}{12\pi^2} 4\pi R_Q^2,
\end{equation}
where $N_q$ is the number of the decay quarks. The temperature at the decay thus reads as
\begin{eqnarray}
T_{\rm D} & = & \left(\frac{90}{4\pi^2 N_{\rm D}}\right)^{1/4}\sqrt{\Gamma_Q^{\rm (q)} M_{\rm P}}
\nonumber \\
& \simeq & 16 \ {\rm MeV} \left(\frac{\zeta}{2.5}\right)^{1/2} 
\left(\frac{N_q}{18}\right)^{1/2} \left(\frac{N_{\rm D}}{10.75}\right)^{-1/4}
\nonumber \\ & & \hspace{10mm} \times
\left(\frac{M_F}{10^6 \ {\rm GeV}}\right)^{1/2} 
\left(\frac{Q}{10^{24}}\right)^{-5/8},
\label{td}
\end{eqnarray}
where $N_{\rm D}$ is the relativistic degrees of freedom at the decay time. Since the decay should take 
place before the big bang nucleosynthesis (BBN), the charge of the gauge-mediation type $Q$ ball
has the upper limit, as shown later.

\section{Baryon and dark matter abundances}
The baryon number of the universe is created by the AD mechanism, spreaded through the decay of the 
gauge-mediation type $Q$ balls. Thus, we obtain
\begin{eqnarray}
\label{Yb}
Y_b & \equiv & \frac{n_b}{s} = 
\frac{3T_{\rm RH}}{4} \left.\frac{n_b}{\rho_{\rm rad}}\right|_{\rm RH} 
= \frac{3T_{\rm RH}}{4} \left.\frac{n_b}{\rho_{\rm inf}}\right|_{\rm rot} \nonumber \\
& \simeq & \frac{9}{8\sqrt{2}}  b \frac{T_{\rm RH} \phi_{\rm rot}^3}{M_F^2 M_{\rm P}^2} \\
& \simeq & \frac{9}{8\sqrt{2}}  b \beta_{\rm G}^{-3/4}\frac{M_F T_{\rm RH}}{M_{\rm P}^2} Q_{\rm G}^{3/4},
\label{Yb_gen}
\end{eqnarray}
where $\varepsilon=1$, $n_b|_{\rm rot} \simeq b m_{\rm eff} \phi_{\rm rot}^2$, 
$m_{\rm eff} = \sqrt{V''}\simeq 2\sqrt{2}M_F^2/\phi_{\rm rot}$, 
$3H_{\rm rot} = m_{\rm eff}$, and Eq.(\ref{Qform}) are used. Since the abundance of the
baryon number is $Y_b \simeq 10^{-10}$, we must have the $Q$ ball with the charge
\begin{eqnarray}
\hspace{-6mm}
Q_{\rm G} & \simeq & 7.8 \times 10^{19} \left(\frac{b}{1/3}\right)^{-4/3}
\left(\frac{\beta_{\rm G}}{6\times 10^{-4}}\right) \nonumber \\
& & \hspace{-10mm} \times \left(\frac{Y_b}{10^{-10}}\right)^{4/3} 
\left(\frac{T_{\rm RH}}{10^4 \, {\rm GeV}}\right)^{-4/3}\left(\frac{M_F}{10^6 \, {\rm GeV}}\right)^{-4/3}.
\label{Qb}
\end{eqnarray}

On the other hand, the new type $Q$ ball is the dark matter. Its abundance is estimated as
\begin{eqnarray}
\label{dm-density}
\frac{\rho_Q}{s} & = & \frac{3T_{\rm RH}}{4} \left.\frac{\rho_Q}{\rho_{\rm inf}}\right|_{\rm rot}
\simeq \frac{9}{4} T_{\rm RH} \left(\frac{\phi_{\rm rot}}{M_{\rm P}}\right)^2 \\
& \simeq & \frac{9}{4} T_{\rm RH} \beta_N^{-1} Q_{\rm N} \left(\frac{m_{3/2}}{M_{\rm P}}\right)^2.
\end{eqnarray}
Here we use $\rho_Q = M_Q n_Q = m_{3/2} m_{\rm eff} \phi_{\rm rot}^2 = m_{3/2}^2 \phi_{\rm rot}^2$,
$\rho_{\rm inf}=3H_{\rm rot}^2M_{\rm P}^2= m_{3/2}^2 M_{\rm P}^2/3$, and Eq.(\ref{form_new}).
Since it is related to the baryon number as $\rho_Q/\rho_b \simeq 5.4$, the charge of the $Q$ ball
should be
\begin{eqnarray}
Q_{\rm N} & \simeq & 2.8\times 10^{22} \left(\frac{Y_b}{10^{-10}}\right)
\left(\frac{\beta_{\rm N}}{0.02}\right)
\nonumber \\ & & \times
\left(\frac{T_{\rm RH}}{10^3\ {\rm GeV}}\right)^{-1}
\left(\frac{m_{3/2}}{\rm GeV}\right)^{-2}.
\label{Qdm}
\end{eqnarray}
In the next section, we consider if these charges of both the gauge-mediation and the new types
are allowed in the $Q$-ball parameter space.

\section{Constraints on the $Q$ ball parameters}
Here, we investigate the constraints on the $Q$-ball parameters, and see how large the charge
could be for both types in order to explain the baryon asymmetry and the dark matter of the universe.

\subsection{Gauge-mediation type}
First let us consider the gauge-mediation type $Q$ ball. The charge of the $Q$ ball
necessary for the baryon number of the universe is expressed as Eq.(\ref{Qb}). There are several
conditions to be satisfied. (a) The $Q$ ball has to decay into nucleons. (b) The decay must complete
before the BBN. (c) The potential should be dominated by $V_{\rm gauge}$. 
In addition to these, (d) $M_F$ has an upper limit (\ref{MFlimit}), and, for simplicity,  (e) we consider the 
case $M_F > T_{\rm rot}$, where the thermal effects can be neglected.

\begin{figure}[ht!]
\includegraphics[width=90mm]{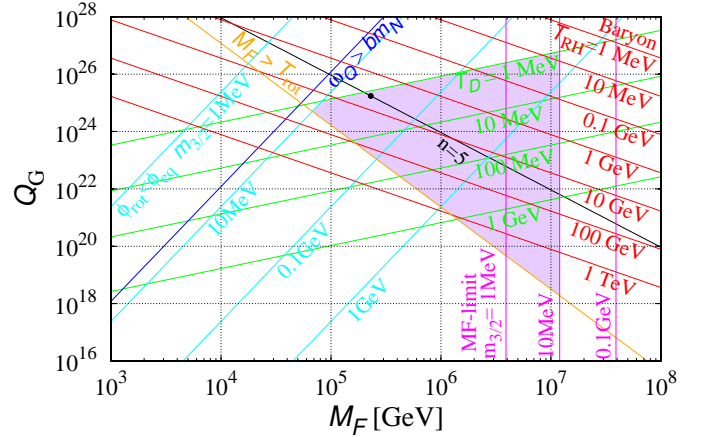}
\caption{Allowed region of the gauge-mediation type $Q$ ball to explain the baryon number of the universe. 
Red lines represent $Y_b=10^{-10}$ for various reheating temperatures. There are several constraints. 
(a) The $Q$ ball is unstable (blue), and (b) decays before the BBN (green). 
(c) $\phi_{\rm rot} < \phi_{\rm eq}$ (cyan). (d) $M_F$ has an upper bound (magenta). 
(e) $M_F > T_{\rm rot}$ (orange). The black line denotes the $n=5$ gauge-mediation type
$Q$ ball with $\lambda_5=0.01$. Shaded region is allowed for $m_{3/2}=10$~MeV.}
\label{fig-gauge}
\end{figure}

In Fig.~\ref{fig-gauge}, we plot the charge of the $Q$ ball which produces the right amount of the baryon number
in red lines for various reheating temperatures. The condition (a) requires 
that $Q_{\rm G} < Q_{\rm D}$, where $Q_{\rm D}$ is defined in Eq.(\ref{QD}), 
displayed by the blue line.

In oder to satisfy the condition (b), $T_{\rm D} > 1$ MeV. We can derive the charge
of the $Q$ ball, which depends on the decay temperature $T_{\rm D}$, from 
Eq.(\ref{td}), as
\begin{eqnarray}
Q_{\rm G} & = & 8.4\times 10^{25} \left(\frac{\zeta}{1.5}\right)^{4/5}\left(\frac{N_q}{18}\right)^{4/5}
\left(\frac{N_{\rm D}}{10.75}\right)^{-2/5} \nonumber \\
& & \times
\left(\frac{T_{\rm D}}{\rm MeV}\right)^{-8/5}
\left(\frac{M_F}{10^6 \ {\rm GeV}}\right)^{4/5}.  
\end{eqnarray}
It is indicated by green lines.

The condition (c) implies that the amplitude of the field at the onset of the rotation should be smaller than
$\phi_{\rm eq}$, where $\phi_{\rm eq}$ is defined by $V_{\rm gauge}(\phi_{\rm eq}) (\simeq M_F^4) = 
V_{\rm grav}(\phi_{\rm eq}) (\simeq\frac{1}{2}m_{3/2}^2\phi_{\rm eq}^2)$, 
and related to the charge $Q_{\rm G}$ in Eq.(\ref{form-gauge}). The constraint, represented by cyan lines is 
written as
\begin{equation}
Q_{\rm G} < 2.4 \times 10^{21} \left(\frac{\beta_{\rm G}}{6\times 10^{-4}}\right)
\left(\frac{M_F}{10^6\, {\rm GeV}}\right)^4
\left(\frac{m_{3/2}}{\rm GeV}\right)^{-4}.
\end{equation}

The upper limit on $M_F$ (d) is shown in magenta lines. The condition (e) comes from the fact that
the scalar potential of Eq.(\ref{pot-gauge}) dominates over the two-loop thermal logarithmic potential at
the onset of the field rotation. Since the inflaton oscillation dominates the energy density of the universe
at that time, $T_{\rm rot} \simeq (M_{\rm P} T_{\rm RH}^2H_{\rm rot})^{1/4}$. Thus, we must have
\begin{equation}
Q_{\rm G} > 1.6\times 10^{30}  \left(\frac{\beta_{\rm G}}{6\times 10^{-4}}\right)
\left(\frac{T_{\rm RH}}{10^4 \, {\rm GeV}}\right)^8
\left(\frac{M_F}{10^6\, {\rm GeV}}\right)^{-12}.
\end{equation}
Eliminating $T_{\rm RH}$ dependence by using Eq.(\ref{Qb}), we obtain
\begin{eqnarray}
Q_{\rm G} & \gtrsim & 2.3\times 10^{21} \left(\frac{\beta_{\rm G}}{6\times 10^{-4}}\right)
\left(\frac{b}{1/3}\right)^{-8/7} 
\nonumber \\ & & \times
\left(\frac{Y_b}{10^{-10}}\right)^{8/7} 
\left(\frac{M_F}{10^6 \, {\rm GeV}}\right)^{-20/7},
\end{eqnarray}
displayed as the orange line. We shade the region for the allowed parameter space for $m_{3/2}=10$~MeV,
for example. Therefore, we see that there is rather large allowed parameter space for the gauge-mediation type.

\subsection{New type}
\begin{figure}[ht!]
\includegraphics[width=90mm]{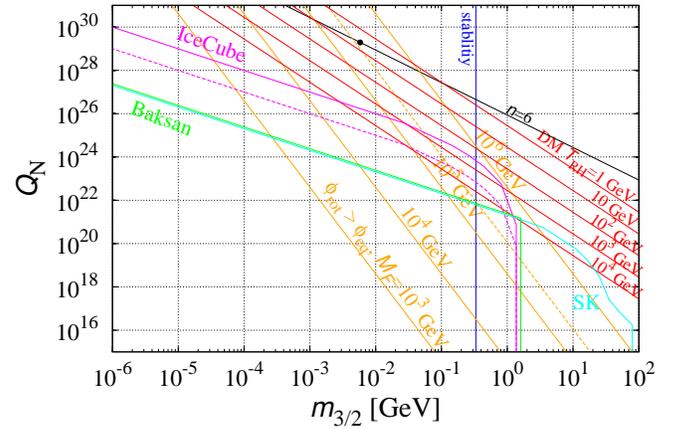}
\caption{Allowed region of the new type $Q$ ball to be the dark matter of the universe. 
Red lines represent $\rho_Q/\rho_b \simeq 5.4$ for various reheating temperatures. 
There are several constraints. (a) The $Q$ ball is stable (blue). 
(b) $\phi_{\rm rot} > \phi_{\rm eq}$ (orange). Observational limits come from the BAKSAN (green), 
the Super Kamiokande (cyan), and the IceCube (magenta).
The black line denotes the $n=6$ new type $Q$ ball.}
\label{fig-new}
\end{figure}

Here we see the constraints on the dark matter new type $Q$ ball. The charge should be Eq.(\ref{Qdm}), shown
in red lines for various reheating temperatures in Fig.~\ref{fig-new}. In this case, we have two theoretical 
conditions: (a) the $Q$ ball must be stable, and (b) the potential is dominated by $V_{\rm grav}$. In addition,
there are some observational constraints. 

The condition (a) is simple: $\omega_Q \simeq m_{3/2} < b m_N$, displayed by the blue line.
Since the boundary $\phi_{\rm eq}$ for the condition (b) is the same as for the previous case, we
obtain the constraint as
\begin{equation}
Q_{\rm N} > 4.0\times 10^{22}  \left(\frac{\beta_{\rm N}}{0.02}\right)
\left(\frac{M_F}{10^6\ {\rm GeV}}\right)^4
\left(\frac{m_{3/2}}{\rm GeV}\right)^{-4},
\end{equation}
where Eq.(\ref{form_new}) is used. It is shown in orange lines.

We also put the observational constraints. Most stringent bounds came from the BAKSAN (green) and the 
Super Kamiokande (cyan) experiments \cite{Arafune, KK3, SK}. Recently, the IceCube experiment 
reported the limits on subrelativistic magnetic monopoles \cite{IC}, which can be applied to
the $Q$-ball search. Since we do not know the detection efficiency $\eta$, we show the constraints
for $\eta=1$ and 0.1 in solid and dashed magenta lines, respectively. We can thus conclude that the new type 
$Q$ balls is indeed a good candidate for the dark matter in the universe, and could be detected directly
by future IceCube-like detectors.

Notice that the dark matter new type $Q$ ball has a good feature that avoids astrophysical constraints by the
neutron star destruction, raised in Ref.~\cite{NS}. As the charge of the $Q$ ball grows by absorbing
the baryon number of the neutron star when it is captured inside the neutron star, the field amplitude 
increases to reach $\phi_{\rm rot}$. At that field value, the $A$ term,
$V_A \simeq m_{3/2} W_{\rm NR}(\phi_{\rm rot})$, becomes comparable to the lifting 
term, $V_{\rm NR}=|W_{\rm NR}(\phi_{\rm rot})|^2$, in the potential. It is this $A$ term that produces
baryon number in the first place in the AD mechanism. Thus, the baryon number violating operators
are effective, so that it leads to the destruction of the $Q$ ball \cite{Konya}. Therefore, It does not affect
the neutron star.

\section{A case study}
We now investigate a concrete model. We consider the multiple directions: one develops into $n=5$ 
gauge-mediation type $Q$ balls and the other forms $n=6$ new type $Q$ balls.
As mentioned earlier, the AD field acquires a large field amplitude during and after inflation before
the rotation starts. This is due to the negative Hubble-induced mass term, $-H_{\rm rot}^2|\Phi|^2$, 
which stems from the SUSY breaking by the inflaton field \cite{DRT}. The actual amplitude of the AD field is 
obtained by the balance of this term and the higher order term. The latter naturally appears from the 
superpotential of the form
\begin{equation}
W_{\rm NR}=\lambda_n \frac{\Phi^n}{nM_{\rm P}^{n-3}},
\end{equation}
where $\lambda_n$ is a constant. In the minimal supersymmetric standard model (MSSM), all the 
flat directions are lifted by $n=4-9$ superpotential, depending on each direction \cite{GKM}. 
In this way, we obtain the field amplitude at the onset of the rotation, and thus the charge of the $Q$ ball,
respectively, as
\begin{equation}
\phi_{\rm 5,rot} \simeq \sqrt{2} \left(\frac{2}{3}\right)^{1/4} \lambda_5^{-1/4} \sqrt{M_F M_{\rm P}},
\label{phi5rot}
\end{equation}
\begin{equation}
Q_{\rm G} \simeq \frac{8}{3}\beta_{\rm G} \lambda_5^{-1} \left(\frac{M_F}{M_{\rm P}}\right)^{-2},
\label{qgform}
\end{equation}
for $n=5$ gauge-mediation type, and
\begin{equation}
\phi_{\rm 6,rot} \simeq \frac{\sqrt{2}}{3^{1/4}} \lambda_6^{-1/4} (m_{3/2} M_{\rm P}^3)^{1/4},
\end{equation}
\begin{equation}
Q_{\rm N} \simeq \frac{2}{\sqrt{3}}\beta_{\rm N} \lambda_6^{-1/2} \left(\frac{m_{3/2}}{M_{\rm P}}\right)^{-3/2},
\label{qnform}
\end{equation}
for $n=6$ new type. 

\begin{figure}[ht!]
\includegraphics[width=90mm]{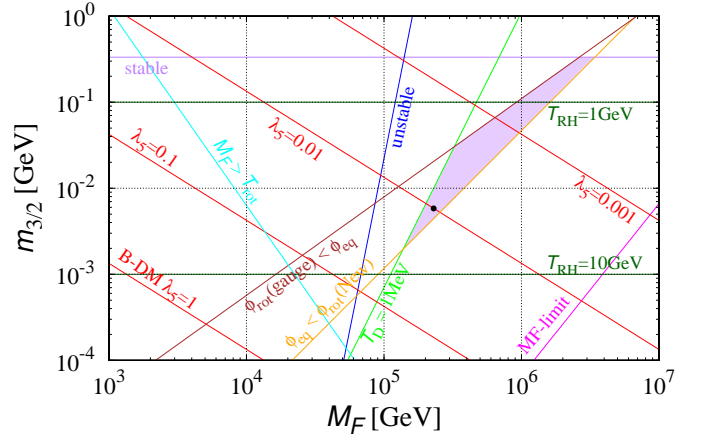}
\caption{Allowed region of the $n=5$ gauge-mediation type and the $n=6$ new type $Q$ balls. 
Red lines represent the right amount of the baryon number and the dark matter for 
various $\lambda_5$. There are several constraints. (a) $n=5$ direction forms the gauge-medition type 
$Q$ ball (brown), which is (b) unstable (blue), and (c) decays before the BBN (green). 
(d) $n=6$ direction becomes the new type $Q$ ball (orange), which is (e) stable (purple).
$T_{\rm RH}$ contour is shown in dark green lines.
Also shown are (f) the upper bound of $M_F$ (magenta),
and (g) $M_F>T_{\rm rot}$ (cyan). Black dot represents one example in the text.}
\label{fig-m32MF}
\end{figure}

Therefore, the baryon abundance is estimated from Eq.(\ref{Yb}) as
\begin{equation}
\frac{\rho_b}{s} = m_N Y_b 
\simeq \left(\frac{3}{2}\right)^{5/4} b \lambda_5^{-3/4}T_{\rm RH}\frac{m_N}{\sqrt{M_F M_{\rm P}}},
\end{equation}
 while the dark matter density is obtained from Eq.(\ref{dm-density}) as
\begin{equation}
\frac{\rho_Q}{s} \simeq \frac{3\sqrt{3}}{2}  \lambda_6^{-1/2} 
T_{\rm RH} \left(\frac{m_{3/2}}{M_{\rm P}}\right)^{1/2}.
\label{rhoDM}
\end{equation}
 Since the ratio $\rho_Q/\rho_b \simeq 5.4$, we must have the relation between $M_F$ and $m_{3/2}$ as
\begin{equation}
m_{3/2} \simeq 1.3\times 10^{-6} \, {\rm GeV}  \lambda_5^{-3/2}  \lambda_6 \left(\frac{b}{1/3}\right)^2
\left(\frac{M_F}{10^6\, {\rm GeV}}\right)^{-1}.
\label{BDM}
\end{equation}
We call it the baryon-dark matter (B-DM) relation. It is shown in red lines for various values
of $\lambda_5$ with $\lambda_6=1$ in Fig.~\ref{fig-m32MF}. Since it is necessary to have 
$Y_b \simeq 10^{-10}$ and $\rho_Q/\rho_b \simeq 5.4$, together with Eq.(\ref{rhoDM}), we obtain 
a contour of the reheating temperature $T_{\rm RH}$ as
\begin{equation}
m_{3/2} \simeq 10^{-3} \, {\rm GeV} \lambda_6 \left(\frac{Y_b}{10^{-10}}\right)^2 
\left(\frac{T_{\rm RH}}{10\, {\rm GeV}}\right)^{-2},
\label{m32Trh}
\end{equation}
displayed by dark green lines.

We can now constrain the B-DM relation by several conditions mentioned earlier already, but 
modify them to this particular situation: (a) The $n=5$ flat direction forms the gauge-mediation type $Q$ ball,
(b) which is unstable, and (c) decays before the BBN. On the other hand, (d) The $n=6$ flat direction forms the 
new type $Q$ ball, (e) which is stable. In addition, (f) $M_F$ should satisfy Eq.(\ref{MFlimit}), and we consider
(g) $M_F > T_{\rm rot}$ for simplicity, where the two-loop thermal potential does not affect the dynamics of
the $n=5$ direction.

Let us first see the conditions (a) -- (c) for the gauge-mediation type. The condition (a) is rephrased as 
$\phi_{\rm 5,rot} < \phi_{\rm eq}$, which leads to
\begin{equation}
m_{3/2} \lesssim 0.72 \, {\rm GeV} \lambda_5^{1/4} \left(\frac{M_F}{10^6\, {\rm GeV}}\right)^{3/2}.
\end{equation}
Since it varies with $\lambda_5$, we can eliminate its dependence using Eq.(\ref{BDM}) to derive
\begin{equation}
m_{3/2} \lesssim 0.11 \, {\rm GeV} \lambda_6^{1/7} \left(\frac{b}{1/3}\right)^{1/7}
\left(\frac{M_F}{10^6\, {\rm GeV}}\right)^{8/7},
\end{equation}
denoted by the brown line. The condition (b) is expressed as $\omega_Q > b m_N$, so that we have
\begin{eqnarray}
M_F & > & 4.4 \times 10^4 \, {\rm GeV} \lambda_5^{-1/6} \left(\frac{b}{1/3}\right)^{2/3} \nonumber \\
& & \times 
\left(\frac{\zeta}{2.5}\right)^{-2/3} \left(\frac{\beta_{\rm G}}{6 \times 10^{-4}}\right)^{1/6},
\end{eqnarray}
where Eq.(\ref{qgform}) is used. Eliminating $\lambda_5$ with Eq.(\ref{BDM}), we get the relation,
shown in the blue line, as
\begin{eqnarray}
m_{3/2} & < & 2.2\times 10^6 \, {\rm GeV} \lambda_6 \left(\frac{b}{1/3}\right)^{-4}
\left(\frac{\zeta}{2.5}\right)^6 \nonumber \\
& & \times \left(\frac{\beta_{\rm G}}{6 \times 10^{-4}}\right)^{-3/2}
\left(\frac{M_F}{10^6\, {\rm GeV}}\right)^8.
\end{eqnarray}
For the condition (c), we must have $T_{\rm D} > 1$~MeV. From Eqs.(\ref{td}) and (\ref{qgform}), we
have 
\begin{eqnarray}
M_F & > & 3.9 \times 10^4 \, {\rm GeV} \lambda_5^{-5/14} \left(\frac{\zeta}{2.5}\right)^{-2/7}
\left(\frac{\beta_{\rm G}}{6 \times 10^{-4}}\right)^{5/14}
\nonumber \\
& & \times 
\left(\frac{N_q}{18}\right)^{-2/7}\left(\frac{N_{\rm d}}{10.75}\right)^{1/7}
\left(\frac{T_{\rm D}}{\rm MeV}\right)^{4/7}.
\end{eqnarray}
With the use of the B-DM relation (\ref{BDM}), 
\begin{eqnarray}
m_{3/2} & < & 1.1\, {\rm GeV} \lambda_6 \left(\frac{b}{1/3}\right)^2
\left(\frac{\zeta}{2.5}\right)^{6/5} 
\nonumber \\ & & \times 
\left(\frac{\beta_{\rm G}}{6 \times 10^{-4}}\right)^{-3/2} 
\left(\frac{N_q}{18}\right)^{6/5}\left(\frac{N_{\rm d}}{10.75}\right)^{-3/5}
\nonumber \\ & & \times 
\left(\frac{T_{\rm D}}{\rm MeV}\right)^{-12/5}
\left(\frac{M_F}{10^6\, {\rm GeV}}\right)^{16/5},
\end{eqnarray}
is obtained, displayed by the green line.

The next is the conditions (d) and (e) for the new type. The condition (d) leads to $\phi_{\rm 6,rot} > \phi_{\rm eq}$,
which results in
\begin{equation}
m_{3/2} \gtrsim 4.7\times 10^{-2} \, {\rm GeV} \lambda_6^{1/5} \left(\frac{M_F}{10^6\, {\rm GeV}}\right)^{8/5},
\end{equation}
shown in the orange line in Fig.~\ref{fig-m32MF}, and the stability condition (e) is nothing but
\begin{equation}
m_{3/2} < 0.33 \, {\rm GeV} \left(\frac{b}{1/3}\right), 
\end{equation}
denoted by the purple line.

There are two more conditions, although they are not restrictive. The condition (f), in the magenta line, reads as
\begin{equation}
m_{3/2} \gtrsim 6.6\times 10^{-5} \, {\rm GeV} g^{-1} \left(\frac{M_F}{10^6\, {\rm GeV}}\right)^2,
\end{equation}
while the condition (g) results in
\begin{eqnarray}
m_{3/2} & \gtrsim & 1.7 10^{-7}\, {\rm GeV} \lambda_6 \left(\frac{Y_b}{10^{-10}}\right)^{12/7}
\nonumber \\ & & \times
\left(\frac{b}{1/3}\right)^{2/7} \left(\frac{M_F}{10^6\, {\rm GeV}}\right)^{-16/7},
\end{eqnarray}
where Eqs.(\ref{phi5rot}), (\ref{m32Trh}), and (\ref{BDM}) are used, shown in the cyan line.

We can now see that it is successful for $M_F=10^5 - 10^6$~GeV, $m_{3/2}=10 - 100$~MeV, 
and $\lambda_5= 10^{-3} - 10^{-2}$ in the model with the $n=5$ gauge-mediation type $Q$ ball
producing the baryon number and the $n=6$ new type $Q$ ball to be the dark matter. 
In order to see the relations among Figs.~\ref{fig-gauge}, \ref{fig-new}, and \ref{fig-m32MF}, 
we also plot the black dots for the particular case of 
$M_F=2.3\times 10^5$~GeV and $m_{3/2}=5.4$~MeV with $\lambda_5=0.01$, for example.
Actually, one good realization of this situation is the flat directions composed of $L$, $u$, $d$ in MSSM.
In this case, four directions are lifted by $n=5$ operators and the last five are lifted by $n=6$ 
operators~\cite{GKM}.

\section{Conclusions}
We have studied the new scenario of the Affleck-Dine $Q$ ball cosmology, in which one flat direction
forms unstable $Q$ balls to create baryon number in the universe by their decay, while the other direction
builds up into stable $Q$ balls to be the dark matter of the universe. Since the gravitino mass is usually
smaller than that of nucleons in the gauge-mediated SUSY breaking, the new type $Q$ ball is always 
stable against the decay into nucleons. At the same time, the gauge-mediation type $Q$ ball with
small enough charge $Q$ can decay into nucleons, releasing all the baryon number they had contained.

We have shown that this scenario works well for wide range of the parameter space: $M_F=10^5 - 10^7$~GeV
and $Q_{\rm G} = 10^{17} - 10^{25}$ for the gauge-mediation type with $m_{3/2} \lesssim$~GeV and
$Q_{\rm N} \gtrsim 10^{25}$ for the new type. We have seen that this dark matter new type $Q$ balls could
be detectable in the (future) IceCube-like experiments.

We have also presented more concrete model: the $n=5$ direction forms gauge-mediation type $Q$ balls, while
the $n=6$ direction becomes the new type $Q$ balls. In this case, we can pin down the model parameters
more accurately. 

In addition, the astrophysical constraints from the neutron star \cite{NS} do not exist. Since the
$Q$ balls form from the rotating AD condensate, and its rotation is due to the $A$ terms in the potential, the same
$A$ terms prevent the $Q$ balls captured by the neutron star from growing up at the certain charge. Thus,
they would not swallow the baryonic charge of the neutron star so much, and are even destroyed
by the effects of the $A$ terms. Therefore, the new type $Q$ ball can safely be the dark matter of the universe.

\section*{Acknowledgments}
The work is supported by Grant-in-Aid for Scientific Research  
23740206 (S.K.), 25400248 (M.K.) and 21111006 (M.K.) 
from the Ministry of Education, Culture, Sports, Science and 
Technology in Japan, and also by World Premier International Research 
Center Initiative (WPI Initiative), MEXT, Japan.

\appendix
\section{$Q$ ball dark matter and baryon number evaporation in a single flat direction}
In this appendix, we show that the scenario with a single flat direction, which forms the dark matter $Q$ balls
with the baryon number evaporated from their surface, explains the dark matter and the baryon asymmetry 
in the universe simultaneously, is now excluded by direct $Q$-ball detection experiments for both the 
gauge-mediation and new types $Q$ balls.

First, let us re-estimate the charge evaporated from the Q ball in the thermal bath, following along the argument
in Ref.~\cite{KK3}. The charge emitting rate is determined by the diffusion in high temperatures and by the 
evaporation in low temperatures. The diffusion rate is written as \cite{BJ}
\begin{equation}
\Gamma_{\rm diff} \equiv \left.\frac{dQ}{dt}\right|_{\rm diff} \simeq - 4 \pi D R_Q \mu_Q T^2,
\end{equation}
where chemical potential is given by $\mu_Q = \omega_Q$, and the diffusion coefficient is $D=A/T$ 
with $A=4-6$. On the other hand, the evaporation rate is expressed as \cite{LaSh}
\begin{equation}
\Gamma_{\rm evap} \equiv \left.\frac{dQ}{dt}\right|_{\rm evap} \simeq 
- 4 \pi \xi (\mu_Q - \mu_{\rm plasma}) T^2 R_Q^2,
\end{equation}
where $\mu_{\rm plasma} \ll \mu_Q$ is the chemical potential of thermal plasma, and 
\begin{equation}
\xi =\left\{\begin{array}{ll}
1 & (T>m_s), \\
\displaystyle{\left(\frac{T}{m_s}\right)^2} & (T < m_s),
\end{array}\right.
\end{equation}
where $m_s$ is the sparticle mass.

Since the $Q$ balls would experience both epochs of the inflaton-oscillation domination (IOD) and 
radiation domination (RD), the transformation from the time $t$ to the temperature $T$ can be written as
\begin{equation}
\frac{dt}{dT} \simeq \left\{\begin{array}{ll}
\displaystyle{-\frac{8}{3}M_{\rm P} T_{\rm RH}^2 T^{-5}} & {\rm (IOD)}, \\
\displaystyle{-\frac{3}{\pi}\left(\frac{10}{N_*}\right)^{1/2} M_{\rm P}  T^{-3}} & {\rm (RD)},
\end{array}\right.
\end{equation}
where $N_*$ is the relativistic degrees of freedom.

\subsection{Gauge-mediation type}
For the gaue-mediation type $Q$ ball, the diffusion and the evaporation rates are respectively given as
\begin{equation}
\Gamma_{\rm diff} \simeq - 4 \pi^2 A T,
\end{equation}
and
\begin{equation}
\Gamma_{\rm evap} \simeq -2\sqrt{2}\pi^2 \xi \zeta^{-1} \frac{T^2}{m(T)} Q^{1/4},
\end{equation}
where 
\begin{equation}
m(T)=\left\{\begin{array}{ll}
T & (T>M_F), \\
M_F & (T < M_F).
\end{array}\right.
\end{equation}
The transition occurs at the temperature
\begin{equation}
T_* = (\sqrt{2} \zeta A)^{1/3} (m_s^2 M_F)^{1/3} Q^{-1/12},
\end{equation}
where $T_*<m_s$ for the stable $Q$ ball ($Q>Q_{\rm D}$).

There are two cases to be considered: (a) $T_*<T_{\rm RH}$ and
(b) $T_{\rm RH} < T_*$. In the former case (a), the evaporated charge is evaluated as
\begin{eqnarray}
& & \Delta Q(T>T_{\rm RH}) \simeq \frac{32}{9} \pi^2A \frac{M_{\rm P}}{T_{\rm RH}}, \\
& & \Delta Q(T_*<T<T_{\rm RH}) \simeq 12 \pi A \sqrt{\frac{10}{N_*}} M_{\rm P} 
\left(\frac{1}{T_*}-\frac{1}{T_{\rm RH}}\right), \nonumber \\ & & \\
& & \Delta Q(T<T_*)  \simeq 3\sqrt{2} \pi \zeta^{-1} \sqrt{\frac{10}{N_*}}\frac{M_{\rm P}T_*^2}{m_s^2M_F} Q^{1/4}, 
\end{eqnarray}
so that the total evaporated charge becomes
\begin{eqnarray}
\hspace{-5mm}
\Delta Q & \simeq& \frac{32}{9} \pi^2A \frac{M_{\rm P}}{T_{\rm RH}}\left(1-\frac{27}{8\pi}\sqrt{\frac{10}{N_*}}\right)
\nonumber \\ & & \hspace{-6mm}
+  2^{5/6} 9\pi A^{2/3}\zeta^{-1/3} \sqrt{\frac{10}{N_*}}\frac{M_{\rm P}}{(m_s^2M_F)^{1/3}} Q^{1/12}.
\label{DeltaQga}
\end{eqnarray}
Here we assume $T_{\rm RH} \ll T_{\rm max}$ and $T_0 \ll T_*$, where $T_{\rm max}$ and $T_0$ is the
maximum temperature after inflation and the present-day temperature, respectively.

On the other hand, in the case (b), we can estimate the evaporated charge as
\begin{eqnarray}
\hspace*{-3mm}
& & \Delta Q(T>T_*) \simeq \frac{32}{9}\pi^2 A \frac{M_{\rm P}T_{\rm RH}^2}{T_*^3}, \\ 
\hspace*{-3mm}
& & \Delta Q(T_{\rm RH}<T<T_*) \simeq \frac{16\sqrt{2}}{3}\pi^2\zeta^{-1}
\frac{M_{\rm P}T_{\rm RH}^2}{m_s^2M_F}Q^{1/4} \log\frac{T_*}{T_{\rm RH}}, \nonumber \\ 
\hspace*{-3mm} & &\\
\hspace*{-3mm}
& & \Delta Q(T<T_{\rm RH}) \simeq 3\sqrt{2}\pi\sqrt{\frac{10}{N_*}}\zeta^{-1}
\frac{M_{\rm P}T_{\rm RH}^2}{m_s^2M_F}Q^{1/4}, 
\end{eqnarray}
so that the total evaporated charge is obtained as
\begin{eqnarray}
\Delta Q \simeq \left( 1 + 3\log\frac{T_*}{T_{\rm RH}} + \frac{27}{16\pi}\sqrt{\frac{10}{N_*}}\right) 
\nonumber \\
\times \frac{16\sqrt{2}}{9}\pi^2\zeta^{-1}\frac{M_{\rm P}T_{\rm RH}^2}{m_s^2M_F}Q^{1/4}.
\label{DeltaQgb}
\end{eqnarray}
The largest evaporated charge is realized for $T_{\rm RH}=T_*$, where Eqs.(\ref{DeltaQga}) and (\ref{DeltaQgb})
become identical, expressed as
\begin{equation}
\Delta Q_{\rm max} \simeq \frac{16\pi^2}{9}2^{5/6} \frac{A^{2/3}}{\zeta^{1/3}} \frac{M_{\rm P}}{(m_s^2 M_F)^{1/3}}
\left(1+\frac{27}{16\pi}\sqrt{\frac{10}{N_*}}\right).
\end{equation}

\begin{figure}[ht!]
\includegraphics[width=90mm]{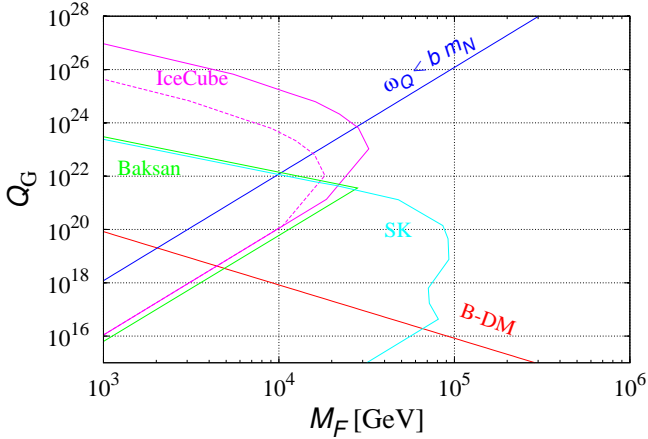}
\caption{Exclusion of the dark matter gauge-mediation type $Q$ ball with the evaporated baryon numbers 
being the baryon asymmetry of the universe. 
Both the dark matter $Q$ ball and the baryon number of the universe are explained on the red line. Blue line
represents the stability condition. Observational limits are displayed for the BAKSAN (green), 
the Super Kamiokande (cyan), and the IceCube (magenta).}
\label{fig-obs-gauge}
\end{figure}

In order to explain both the baryon number and the dark matter densities in the universe, 
their ratio should be
\begin{equation}
\frac{\rho_Q}{\rho_b} = \frac{M_Q n_Q}{m_N n_b} = \frac{M_Q}{m_N \varepsilon b \Delta Q} \simeq 5.4.
\end{equation}
Therefore, the charge $Q_{\rm G}$ should hold
\begin{eqnarray}
Q_{\rm G} & = & 8.3\times 10^{13} \varepsilon^{3/2} \left(\frac{A}{4}\right) \left(\frac{\zeta}{2.5}\right)^{-2}
\left(\frac{b}{1/3}\right)^{-3/2} 
\nonumber \\ & & \times
\left(\frac{m_s}{\rm TeV}\right)^{-1} \left(\frac{M_F}{10^6 \, {\rm GeV}}\right)^{-2},
\end{eqnarray}
where $N_*=200$ is used, shown in the red line in Fig.~\ref{fig-obs-gauge}.

We also displayed observational limits in the figure. The green and cyan lines come from the BAKSAN and
the Super Kamiokande experiments. Together with the stability limit (the blue line), they already exclude
this scenario. Also shown is the IceCube results (in magenta lines).

\subsection{New type}
For the new type $Q$ ball, the diffusion and the evaporation rates are written as
\begin{equation}
\Gamma_{\rm diff} \simeq - 4\pi A|K|^{-1/2} T,
\end{equation}
and
\begin{equation}
\Gamma_{\rm evap} \simeq - 4\pi \xi  |K|^{-1} \frac{T^2}{m_s},
\end{equation}
respectively. The diffusion is the bottleneck of the process at high temperatures, while the evaporation 
determines the process at low temperatures. The transition takes place at the temperature,
\begin{equation}
T_* \simeq \left( A |K|^{1/2} \right)^{1/3} \left(m_s^2 m_{3/2} \right)^{1/3}.
\end{equation}

\begin{figure}[ht!]
\includegraphics[width=90mm]{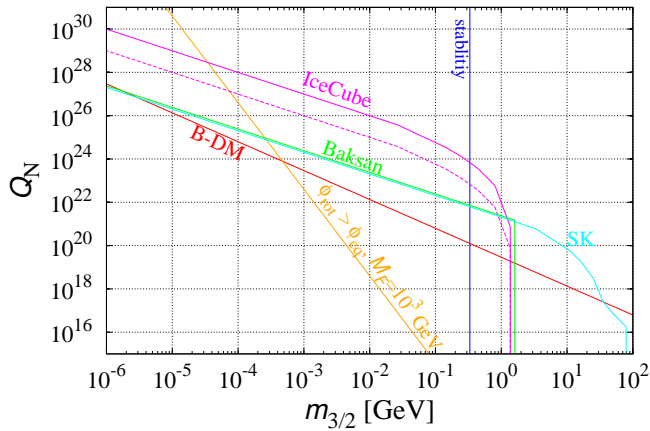}
\caption{Exclusion of the dark matter new type $Q$ ball with the evaporated baryon numbers 
being the baryon asymmetry of the universe. 
Both the dark matter $Q$ ball and the baryon number of the universe are explained on the red line. Blue line
represents the stability condition, and $\phi_{\rm rot}>\phi_{\rm eq}$ is shown in the orange line. 
Observational limits are displayed for the BAKSAN (green), 
the Super Kamiokande (cyan), and the IceCube (magenta).}
\label{fig-obs-new}
\end{figure}

The evaporated charge is calculated for case (a) $T_* < T_{\rm RH}$, as
\begin{eqnarray}
\hspace{-8mm}
& & \Delta Q(T>T_{\rm RH}) \simeq \frac{32}{9} \pi A |K|^{-1/2} \frac{M_{\rm P}}{T_{\rm RH}}, \\
\hspace{-8mm}
& & \Delta Q(T_*<T<T_{\rm RH}) \nonumber \\
\hspace{-8mm}
& & \hspace{13mm} \simeq 12  A |K|^{-1/2} \sqrt{\frac{10}{N_*}} M_{\rm P}
\left(\frac{1}{T_*}-\frac{1}{T_{\rm RH}}\right), \\
\hspace{-8mm}
& & \Delta Q(T<T_*) \simeq 6 |K|^{-1} \sqrt{\frac{10}{N_*}} \frac{M_{\rm P}T_*^2}{m_s^2m_{3/2}}, 
\end{eqnarray}
so that the total evaporated charge becomes
\begin{eqnarray}
\Delta Q & \simeq& \frac{32}{9} \pi A |K|^{-1/2} \frac{M_{\rm P}}{T_{\rm RH}}
\left(1-\frac{27}{8\pi}\sqrt{\frac{10}{N_*}}\right)
\nonumber \\
& + & 18 A^{2/3} |K|^{-2/3} \sqrt{\frac{10}{N_*}}\frac{M_{\rm P}}{(m_s^2m_{3/2})^{1/3}}.
\label{DeltaQa}
\end{eqnarray}
For the case (b) $T_{\rm RH} < T_*$, we have
\begin{eqnarray}
\hspace{-2mm}
& & \Delta Q(T>T_*) \simeq \frac{32}{9}\pi |K|^{-1}\frac{M_{\rm P}T_{\rm RH}^2}{T_*^3}, \\ 
\hspace{-2mm}
& & \Delta Q(T_{\rm RH}<T<T_*) \simeq \frac{32}{3}\pi |K|^{-1}
\frac{M_{\rm P}T_{\rm RH}^2}{m_s^2m_{3/2}}Q^{1/4} \log\frac{T_*}{T_{\rm RH}}, \nonumber \\ 
\hspace{-2mm} & &\\
\hspace{-2mm}
& & \Delta Q(T<T_{\rm RH}) \simeq 6 |K|^{-1} \sqrt{\frac{10}{N_*}}
\frac{M_{\rm P}T_{\rm RH}^2}{m_s^2M_F}, 
\end{eqnarray}
so that the total evaporated charge is obtained as
\begin{eqnarray}
\Delta Q \simeq \left( 1 + 3 \log\frac{T_*}{T_{\rm RH}} + \frac{27}{16\pi}\sqrt{\frac{10}{N_*}}\right) 
\nonumber \\
\times \frac{32}{9}\pi |K|^{-1} \frac{M_{\rm P}T_{\rm RH}^2}{m_s^2m_{3/2}}.
\label{DeltaQb}
\end{eqnarray}
The largest evaporated charge is achieved for $T_{\rm RH} = T_*$, where the case (a) and (b) becomes the same
as
\begin{equation}
\Delta Q_{\rm max} \simeq \frac{32\pi}{9}\frac{A^{2/3}}{|K|^{2/3}} \frac{M_{\rm P}}{(m_s^2 m_{3/2})^{1/3}}
\left(1+\frac{27}{16\pi}\sqrt{\frac{10}{N_*}}\right).
\end{equation}
Therefore, in order to explain both baryon asymmetry and the dark matter of the universe, the charge of the 
new type $Q$ ball should be
\begin{eqnarray}
Q_{\rm N} & \simeq & 2.9\times 10^{19} \varepsilon \left(\frac{A}{4}\right)^{2/3} \left(\frac{|K|}{0.01}\right)^{-2/3}
\nonumber \\ & & \times
\left(\frac{m_s}{\rm TeV}\right)^{-2/3} \left(\frac{m_{3/2}}{\rm GeV}\right)^{-4/3},
\end{eqnarray}
shown in the red line in Fig.~\ref{fig-obs-new}.

We also plot the stability condition, $\omega_Q < b m_N$, where the left hand side of  the blue line is allowed, 
and the constraint on the amplitude of the field at the onset of the rotaion: $\phi_{\rm rot} > \phi_{\rm eq}$ with
$M_F=10^3$~GeV, where above the orange line is allowed. 
The BAKSAN (green) and the Super Kamiokande (cyan) experiments exclude all of the region
in the parameter space. The IceCube (magenta) results are also displayed. Therefore, we conclude that
the new type $Q$ ball cannot be the dark matter with the charge evaporated being the baryon number 
of the universe.



\end{document}